\documentclass{aa}
\usepackage{graphicx}
\usepackage{amsmath}
\usepackage{natbib}
\usepackage{epsfig}
\usepackage{subfigure}
\usepackage{multirow}
\usepackage{txfonts}
\begin{document}

   \title{Amplitudes and lifetimes of solar-like oscillations  \\ observed by CoRoT\thanks{The CoRoT space mission, launched on 2006 December 27, was developed and is operated by the CNES, with participation of the Science Programs of ESA, ESA's RSSD, Austria, Belgium, Brazil, Germany and Spain.}}
   \subtitle{Red-giant versus main-sequence stars }

 \titlerunning{Oscillation amplitude and damping in red giant and main-sequence stars}


   \author{F. Baudin\inst{1} \and C. Barban\inst{2} \and K. Belkacem\inst{3,2}
   	\and S. Hekker\inst{4,5,6}  \and T. Morel\inst{3} \and \\
	R. Samadi\inst{2}\and
	O. Benomar\inst{1} \and M.-J. Goupil\inst{2} \and
	F. Carrier\inst{4} \and J. Ballot\inst{7} \and \\
          S. Deheuvels\inst{2} \and J. De Ridder\inst{4} \and
          A.P. Hatzes\inst{8} \and T. Kallinger\inst{9,10}
          \and W.W. Weiss\inst{9}}

   	\institute{Institut d'Astrophysique Spatiale, CNRS, Universit\'e Paris XI,
   	91405 Orsay Cedex, France
   	\and LESIA, Universit\'e Pierre et Marie Curie, Universit\'e
	Denis Diderot, Obs. de Paris, 92195 Meudon Cedex, France
	\and Institut d'Astrophysique et de G\'eophysique, Universit\'e de
	Li\`ege, All\'ee du 6 Ao\^ut 17-B 4000 Li\`ege, Belgium	
	\and Instituut voor Sterrenkunde, K.U. Leuven, Celestijnenlaan 200D, 	3001 Leuven, Belgium
	\and School of Physics and Astronomy, University of Birmingham,
	 Edgbaston, Birmingham B15 2TT, UK
	 \and Astronomical Institute 'Anton Pannekoek', University of Amsterdam, Science Park 904, 1098 XH Amsterdam, The Netherlands
	\and Institut de Recherche en Astrophysique et Plan\'etologie, CNRS, Universit\'e de Toulouse, 31400 Toulouse, France
	\and Th{\"u}ringer Landessternwarte, D-07778 Tautenburg, Germany
	\and University of Vienna, Institute for Astronomy,T{\"u}rkenschanzstrasse 17, A-1180 Vienna, Austria
	\and Department of Physics and Astronomy, University of British Columbia, 6224 Agricultural Road, Vancouver, BC V6T 1Z1, Canada
             }

   \date{Received / accepted }

 
  \abstract
  {The advent of space-borne missions such as CoRoT or \textit{Kepler} providing photometric data has brought new possibilities for asteroseismology across the H-R diagram. Solar-like oscillations are now observed in many stars, including red giants and main-sequence stars.}
   {Based on several hundred identified pulsating red giants, we aim to characterize their oscillation amplitudes and widths. These observables are compared with those of main-sequence stars in order to test trends and scaling laws for these parameters for main-sequence stars and red giants.}
   {An automated fitting procedure is used to analyze several hundred Fourier spectra.  For each star, a modeled spectrum is fitted to the observed oscillation spectrum, and mode parameters are derived.}
   {Amplitudes and widths of red-giant solar-like oscillations are estimated for several hundred modes of oscillation. Amplitudes are relatively high (several hundred ppm) and widths relatively small (very few tenths of a $\mu$Hz).}
   {Widths measured in main-sequence stars show a different variation with the effective temperature from red giants. A single scaling law is derived for mode amplitudes of red giants and main-sequence stars versus their  luminosity to mass ratio. However, our results suggest that two regimes may also be compatible with the observations.}

   \maketitle
%

\section{Introduction}
\label{sec:intro}

Probing stellar interiors through measurements of the characteristics of their eigen-oscillations has been limited to few stars. With ground-based, spectroscopic observations and the launch of space experiments of high photometric precision such as MOST, CoRoT and \textit{Kepler}, stellar seismology promises rich harvests. This is true for different types of stars: individual eigen-modes of some main-sequence stars have been observed photometrically \citep[e.g.][]{Michel08}, but in the case of red giants, oscillations have been detected in several hundred stars thanks to CoRoT observations \citep{deRidder09,Hekker09}, whereas previously this kind of oscillations had been detected in only about a dozen red giants \citep[see, among others,][]{Frandsen02,Barban07,Tarrant07,Zechmeister08}. More recently, the launch of the \textit{Kepler} mission provided a larger number of oscillating red giants than CoRoT and allowed some analyses of their global seismic characteristics \citep{Bedding10,Huber10,Kallinger10b}, while the analysis  of red giants detected by CoRoT continue \citep{Mosser10,Kallinger10a,Hekker10}. Because of their size, red giants present oscillations with relatively low frequencies (roughly from 10 to 100\,$\mu$Hz).
The length and the continuity of observations allowed by space-borne experiments is thus very important for the detection of these oscillations. The continuity and stability of the time series yield Fourier spectra unperturbed at low frequency, where the acoustic oscillations of red-giant stars are expected.

Our knowledge of stellar oscillations can now be tested for evolved stars.  Oscillating red-giant stars can of course be individually analyzed \citep[e.g.][]{Carrier10}, but the first consequence of their large number is the need for an analysis method sufficiently automated to allow the exploitation of these samples. As another consequence, the large samples provide a global view of the analyzed objects, for example, using scaling relations between mode parameters and global stellar parameters that provide a general view across the whole H-R diagram. Some scaling laws have been derived \citep[e.g.][]{Kjeldsen95,Chaplin08,Chaplin09,Hekker09,Stello09} using ground-based data and more recently space missions.
A recent and convincing example was the use of the large frequency separation distribution and the frequency of the maximum in the power spectrum, with which the population of CoRoT red giants was identified to be mainly stars of the red clump \citep[He-core burning stars, see][for details]{Miglio09}.

These scaling relations provide in particular new insights into the stochastic excitation and damping of the modes. The latter is still poorly understood even for solar $p$-modes. \citet{Goldreich91} have shown that both radiative losses and turbulent viscosity play a major role in mode damping. In contrast, \citet{Gough80} and \citet{Balmforth92} found that the damping is dominated by the modulation of turbulent pressure, whereas the results of \citet{MAD06} suggest that the perturbation of the convective heat flux is dominant. One can then expect that each contribution exhibits a different relation with fundamental parameters, hence one can seek scaling relations to help identify the dominant physical mechanism involved in mode damping. A first scaling law between the width and the effective temperature $T_{\rm eff}$ of the star has been proposed by \citet{Chaplin09}.

The situation is less critical for mode excitation because the underlying mechanism responsible for mode driving is identified as Reynolds stresses, which are induced by turbulent convection \cite[e.g.][]{Balmforth92,Stein01,Belkacem06}. Scaling laws have been proposed \citep[e.g.][]{Kjeldsen95,Houdek99,Samadi07} with mode amplitudes (in velocity) varying as $(L/M)^s$, where $L$ is the luminosity and $M$ the mass of the star. The issue here is the derivation of the value of the exponent $s$ for the intensity (and then velocity) fluctuations. However, a difficulty arises from the velocity-intensity relation, which is still poorly understood and strongly depends on the non-adiabatic effects in the uppermost layers of the stars that possess convective envelopes.\\

Because mode amplitudes vary as $(L/M)^s$, red giants present a favorable case for oscillation detection as long as the instrument used provides long and continuous observations with low noise at low frequency, which is the case for instruments such as MOST, CoRoT and \textit{Kepler}.
\citet{Hekker09} have used automated analyses to determine the global characteristics of the acoustic spectrum of red giants (mainly the global maximum amplitude $A_{\rm max}$ and its frequency $\nu_{\rm max}$). However, the characteristics of individual modes, for example their width and height, are important to understand mode excitation and damping. Our objective here is to investigate, using CoRoT data, the distribution of amplitude and linewidth, and their dependence on global stellar parameters such as the effective temperature or the luminosity-mass ratio. We present here the automated method used to extract the characteristics of modes (in Section~\ref{sec:method}) and the results obtained for the red giants (in Section~\ref{sec:resRG}). In Section~\ref{sec:comp} these results are compared with those for main-sequence stars.

\section{Data and fitting procedure}
\label{sec:method}

\subsection{Data}

The data\footnote{Data available at {\tt idoc-corot.ias.u-psud.fr/}} used here are the CoRoT data for the observation run starting on 2007 May 16, and ending on 2007 October 10 in the direction of the center of the galaxy (corresponding to the ``Long Run center 01'', or LRc01). The observation length is 142\,days (with a frequency resolution of 0.08\,$\mu$Hz). The stars observed are part of the program aiming at the detection of exo-planets \citep[observed in the ``exo-field'' of the instrument, see][for more details]{Auvergne09}.

\subsection{Fitting procedure}

Our objective is to determine the mode frequencies, widths, and amplitudes from the spectra of more than 300 oscillating red giants reported by \citet{deRidder09} and \citet{Hekker09}. The method used is the same as in many other analyses of oscillating stars: the fitting of a model spectrum to the observed one. The model spectrum consists of a sum of Lorentzian-shaped peaks superimposed on a background (due to noise, granulation, activity...), but considered flat on the narrow frequency interval of interest. A more sophisticated modeling of the noise (varying with frequency for example) can be used, but this adds more free parameters to the model. This is known to increase the instability of the fitting process so we have used a simpler model. We nevertheless performed some tests on a small number of targets, which showed that the differences between the two models lead to measures differing by quantities smaller or, in the worst cases, comparable to the error bars.\\
The best fit to the observations is determined by the maximisation of  a likelihood estimator (MLE). This method has been succesfully applied to the Sun \citep[see for example][for an early application]{Toutain92} and more recently to other stars \citep[see for example][for HD49933]{Appourchaux08}. However, we applied this method in an automated manner to our large sample of stars.\\

The procedure consists of several steps, the first using the results of \citet{Hekker09}: $\nu_{\rm max}$, the oscillation frequency of the maximum height of the heavily smoothed spectrum, to which a Gaussian was fitted. Because we aim here to automatically fit the highest peak of the spectrum, and because red giants show oscillations in a relatively narrow frequency range, we considered a window of only 10\,$\mu$Hz centered on $\nu_{\rm max}$. Then, we computed the mean level of noise in this window. The peaks with a height below a threshold of five times the mean level of the noise (which roughly corresponds to a confidence level of 90\%) were not considered.\\

An important step in the analysis is the choice of the input (or ``guess'') parameters used in the fitting process. For each peak above the threshold, the input frequency is that of the highest point of the peak. The input height is a fraction (0.5) of the maximum height of the peak, which can be expected from the $\chi^2$ distribution of the power spectrum that allows a considerable excursion above the mean value.

\begin{figure}
\includegraphics[width=8cm]{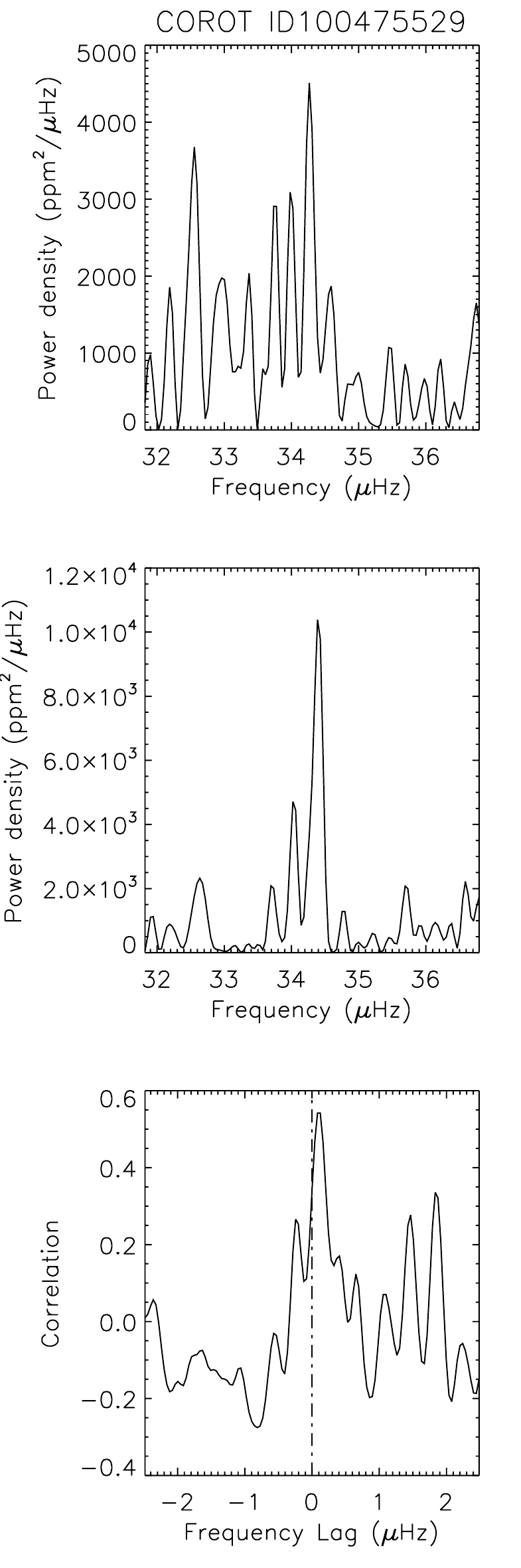}
\caption{Spectra of the two halves of the total time series and their correlation versus the frequency lag. Because the correlation is not maximum for a zero frequency shift, the mode is considered as not coherent during the whole duration of observation.}
\label{fig:2spec}
\end{figure}

\begin{figure}
\includegraphics[angle=90,height=7.5cm]{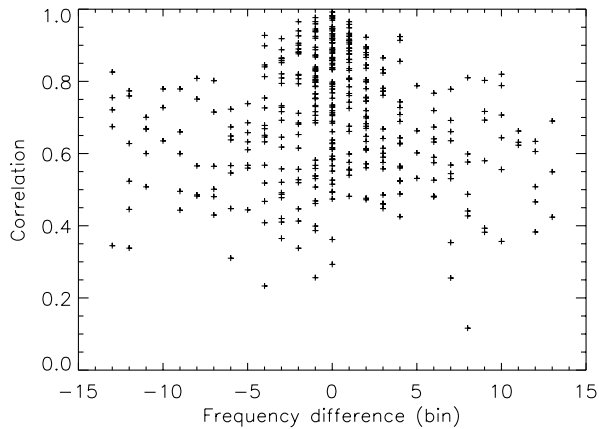}
\caption{Values and shift of the correlation between the spectra computed from two half series for the stars of the sample.}
\label{fig:correl}
\end{figure}

The width is more complex.
Many stars show very narrow peaks, close to the frequency resolution of the spectrum. Hence, we considered two cases: a width smaller (unresolved mode) or larger (resolved mode) than the frequency resolution of the spectrum. If the mode is resolved, the fitted model is a Lorentzian, but an unresolved mode has to be fitted with a sinc function. An automated test was performed to decide which model to use. This test is based on the expected behavior of a long-lived (unresolved) mode compared with a resolved mode: a mode with a lifetime of the order of the observation period is supposed to have its maximum at exactly the same frequency from one observation to another, whereas a short-lived and stochastically excited mode shows a maximum location that may change in frequency (within the mode width $\Gamma=1/\pi \tau$ where $\tau$ is its lifetime) from one observation to another.
Each time series analyzed here was divided into two halves, and the spectra of both sub-series were cross-correlated (see Fig.\,\ref{fig:2spec}). If the highest point of the correlation is obtained for a zero or one frequency-bin shift between the two spectra and shows a high value of correlation, the mode is considered to correspond to a long-lived mode. We then fitted a sinc model with a width equivalent to the frequency resolution. If the highest correlation between the two spectra is obtained for a frequency shift equal to  or larger than twice the frequency resolution or if it has a low value, the model used for the fit is a Lorentzian. Figure\,\ref{fig:correl} shows the correlation values and shifts for the stars of the sample. However, it should be noted that this test can suffer from a low signal-to-noise ratio that causes spiky structures (and then considering a long-lived mode as wider than the frequency resolution). A mode whose coherence time is between the observation period and half its value can also be considered as short-lived. This test was performed with simulated data and has worked properly in about 75\% of the simulations.\\

Several combinations of initial guesses for the mode height and width were tested. It appeared that too low an input value for the width can bias the overall results. Thus, this parameter was chosen to be sufficiently large. Input values for height and background have less influence. Width and height are highly anti-correlated: if one of these parameters is over- or under-estimated, the other will be inversely estimated. Thus, the estimate of the amplitude (the total power), which is computed from the product of width and height, is generally more robust.\\

It must be kept in mind that the results of such an automated procedure may be less accurate for some individual stars among the several hundreds of  analyzed stars. This is inherent to the method, which does not aim at precise results for an individual star \citep[as done by][for instance]{Carrier10} but instead aims at a general estimate to seek trends of the results in a large sample.

\section{Width and amplitude of red-giant oscillations}
\label{sec:resRG}

\subsection{Results}

For each star, an ensemble of parameters is estimated from the fitted model spectrum: the frequency ($\nu$), the height ($H$), and the width ($\Gamma$) of the highest modes in the  frequency range of interest (10\,$\mu$Hz centered on $\nu_{\rm max}$).
We emphasize that in the present work the amplitude of the modes is determined from individual mode measurements and not from a heavily smoothed spectrum where only the mode envelope is visible.

\begin{figure*}
\includegraphics[angle=90,height=7.5cm]{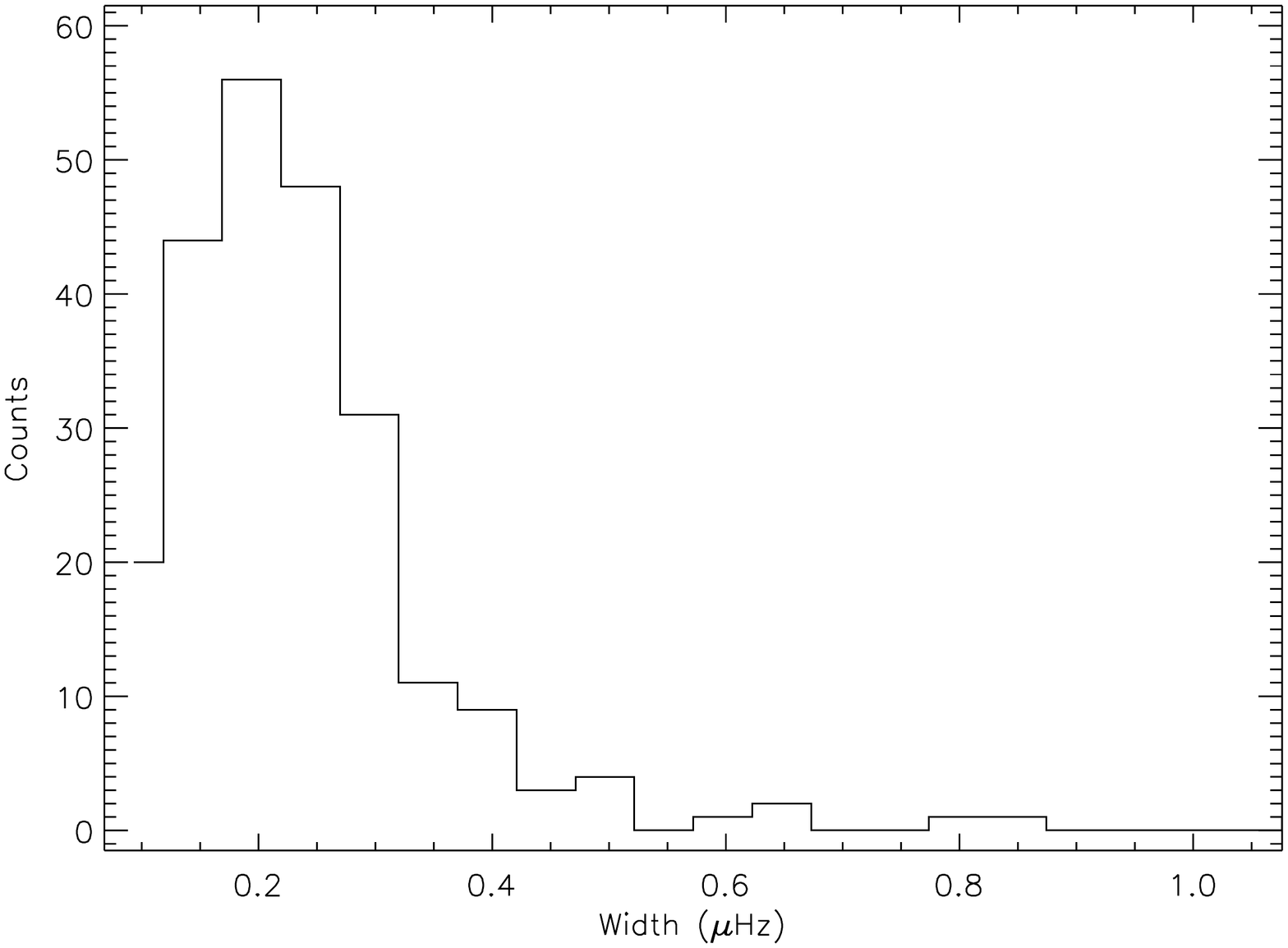}
\includegraphics[angle=90,height=7.5cm]{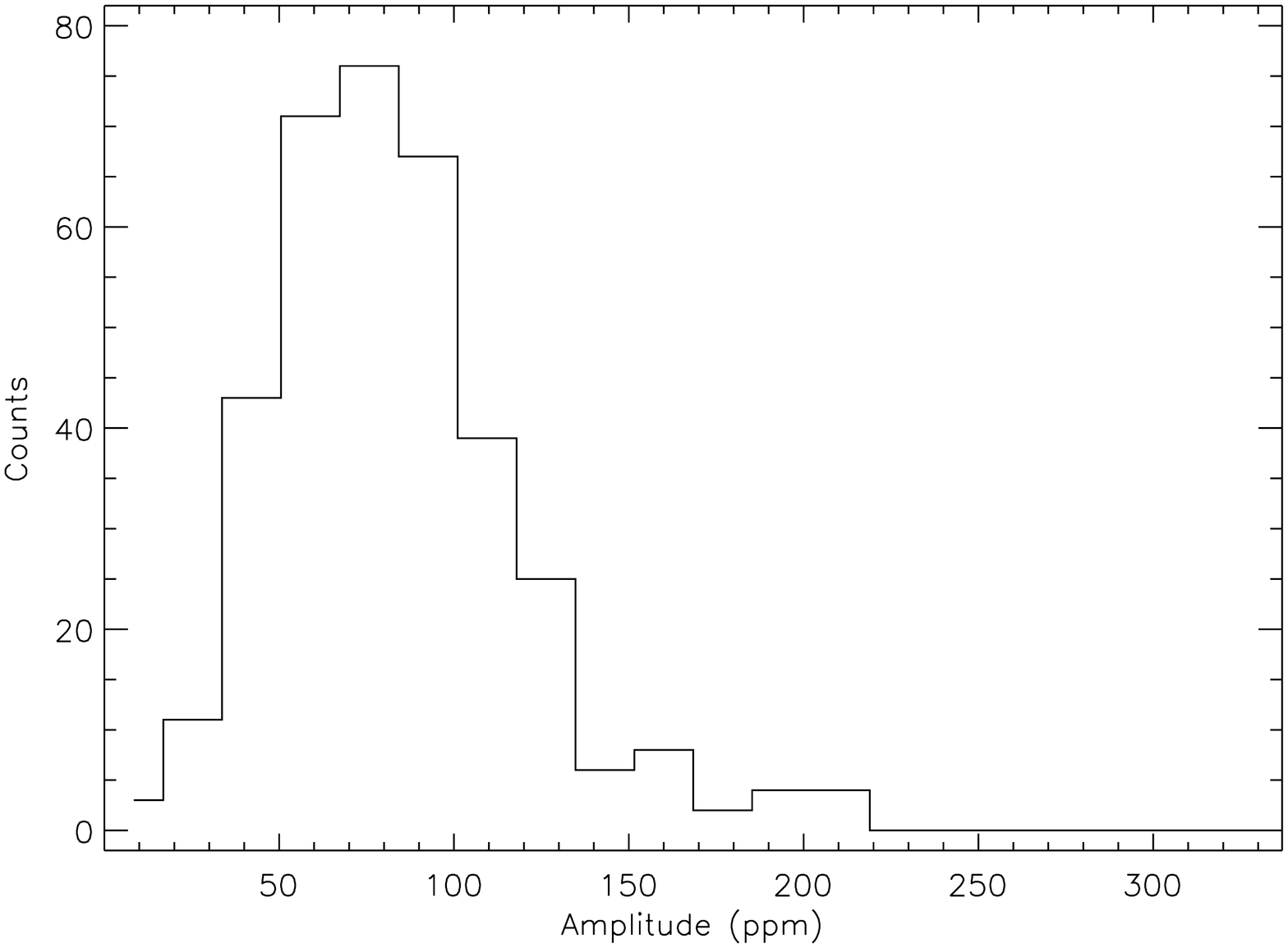}
\caption{Left: distribution of the measured mode linewidths for the spectra fitted with Lorentzian profiles; right: distribution of the measured mode amplitudes, with sinc and Lorentzian fitting.}
\label{fig:histolargamp}
\end{figure*}

Among a total of 398 stars, 35 were not fitted because no peak reached the detection threshold (five times the mean noise level) or had obviously wrong fit results (frequency out of the fitted interval, width of several $\mu$Hz, etc...). Among the 363 remaining spectra, based on the correlation of the two spectra made of half series described in Sect.\,\ref{sec:method}, we fitted 128 (35\%) stars with sinc function profiles and 235 (65\%) with Lorentzian profiles. As mentioned earlier, some of these spectra may have been fitted with a Lorentzian because of a low signal-to-noise ratio, which can be considered as the signature of a short-lived mode. Nevertheless, the fraction of modes fitted with a sinc profile indicates that an important fraction of the analyzed stars have modes with a long lifetime (of the order of or longer than 45 days). When this lifetime is shorter and thus implies the use of a Lorentzian profile, the distribution of measured widths is concentrated at very small widths, close to the frequency resolution (Fig.\,\ref{fig:histolargamp}). This confirms what we saw by eye from an inspection of the spectra.

\begin{figure}
\includegraphics[angle=90,height=7.5cm]{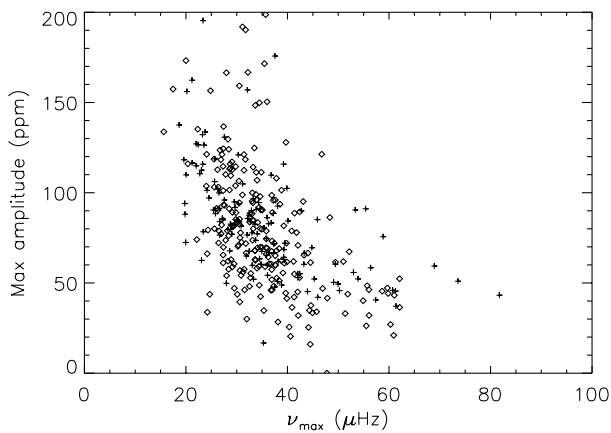}
\caption{Distribution of the measured mode amplitudes, with sinc (crosses) and Lorentzian (diamonds) fitting, versus $\nu_{\rm max}$.}
\label{fig:amp_numax}
\end{figure}

The root-mean-squared amplitude is defined as
$A=\sqrt{\pi H \Gamma}$  in the case of a Lorentzian profile \citep[see for example][]{Baudin05}, which is the area of the profile; or
$A=\sqrt{2H\delta\nu}$ if the lifetime of the oscillation is longer than the observation.
We are here interested in the maximum amplitude: it is taken as the mean of the three (or less if less than three modes were found above the detection level) highest amplitudes. The distribution of the amplitudes in both fitting cases is shown in Fig.\,\ref{fig:histolargamp}. The distribution of these same amplitudes versus $\nu_{\rm max}$ is displayed in Fig.\,\ref{fig:amp_numax}, showing a clearly decreasing trend with increasing $\nu_{\rm max}$.

\subsection{Uncertainties and possible biases}

A possible bias encountered in the mode width estimate is obviously the frequency resolution of the spectra analyzed here: $\delta\nu=0.08\,\mu$Hz.
Thus, the results concerning the width are limited by this resolution as illustrated by the fraction of modes fitted with a sinc profile. However, another important fraction is fitted with Lorentzian profile that yields a distribution of widths that reach several times the value of $\delta \nu$. In addition, one can consider the independent results of \citet{Carrier10}, who analyzed another red giant star observed with CoRoT for more than 140 days (resulting in a frequency resolution of 0.08\,$\mu$Hz, similar to the spectra used in this work). They fitted a Lorentzian profile to the individual observed peaks and found an averaged width of 0.25$\pm$0.06\,$\mu$Hz. This measurement can be compared with the present measurements on Fig.\,\ref{fig:larg}, just above the linear fit to the observed red giant widths. Moreover, \citet{Hekker10} took the opportunity of four common targets in two CoRoT runs to obtain series as long as 494 days, but with a gap of 205 days. The widths obtained for the four stars have an average of 0.19\,$\mu$Hz, very similar to the values we found (see Fig.\,\ref{fig:histolargamp}). The sample is much smaller however, and presents a high dispersion ($\sigma=0.19\mu$Hz).
In addition, and as discussed below, the present results are compatible with recent theoretical predictions \citep{Dupret09}: owing to their inertia, modes can be as narrow as we observe.

Concerning the mode amplitudes, a source of uncertainty is the influence of the effective temperature combined with the instrument sensitivity on the measured oscillation, as described by \citet{Michel09}, who derived a bolometric correction for amplitudes. This correction induces an absolute decrease of about 20\% around $T_{\rm eff}$=4500\,K. However, the relative correction is of the order of a few percent around this temperature. Thus, it will be applied below when we compare red giants with other, hotter, stars. The applied correction is valid for radial modes, but it should be noted that the degree of the fitted peaks of our sample is not known. Because of geometric visibility, modes with a degree of $l$=1 can have a greater height than $l$=0 modes. Thus, a source of uncertainty is the degree of the fitted modes (supposed to be zero) and the adequacy of the bolometric correction. The first can induce an overestimate of about 20\% in amplitude, the second, an underestimate  of at most 10\%.

\subsection{Stellar parameters}
\label{sec:stelparam}

The discussion of the results presented here requires a certain knowledge of the characteristics of the analyzed stars. Scaling laws for seismic parameters will be discussed in the following sections to provide insights into the stellar processes we are interested in: the driving and the damping of the oscillations. The scaling laws used here rely mainly on two parameters: the effective temperature and the luminosity-mass ratio.

The effective temperature, $T_{\rm eff}$, can be derived from photometric broad-band measurements at different wavelengths. Data for the stars analyzed here were derived from the 2MASS data extracted from the Exo-Dat database \citep{Deleuil09} using the relations of \citet{Alonso99}. Only two stars are not in the 2MASS point-source catalog. The magnitudes for  the remaining stars have a high precision.
The 2MASS data were transformed into values in the TCS sytem via a transformation into CIT magnitudes \citep{Alonso98,Carpenter01}).
Solar metallicity was assumed to compute the $T_{\rm eff}$ values. Taking the typical iron abundance of the red giants in this field inferred from population synthesis \citep[[Fe/H{]} $\sim -\,$0.15 dex,][]{Miglio09} would only increase $T_{\rm eff}$($J-H$) by 15 K, while $T_{\rm eff}$($J-K_S$) would remain unaffected.

\begin{figure}
\centering
\includegraphics[width=8.5cm]{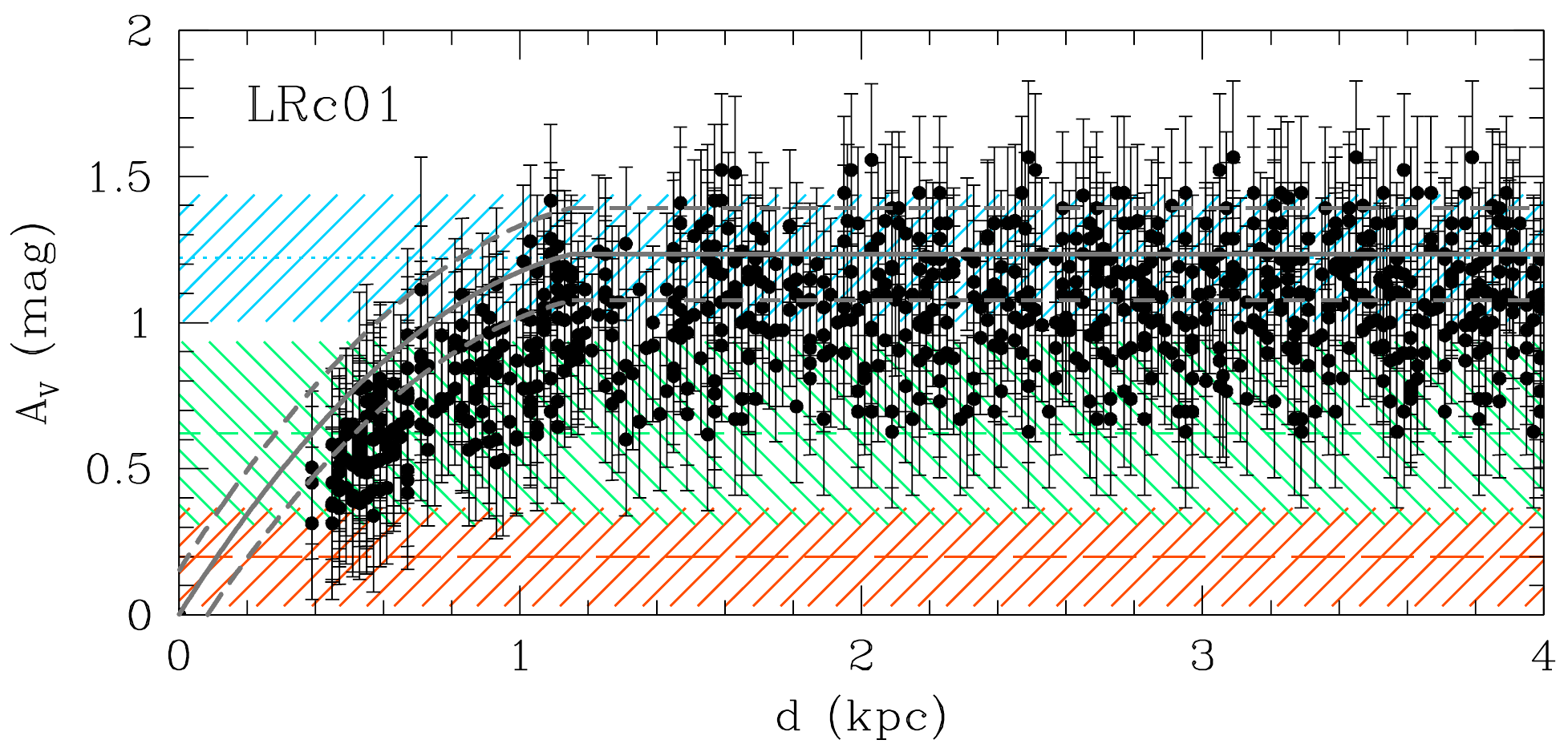}
\caption{Mean extinction values in the Johnson $V$ filter, $A_V$, from 2-D extinction maps based on DSS star counts \citep[][red long-dashed line]{Dobashi05}, 2MASS star counts \citep[][green short-dashed line]{Rowles09}, and COBE/DIRBE with IRAS/ISSA data \citep[][blue dotted line]{Schlegel98}.
The distance-dependent calibration of \citet{Arenou92} based on optical photometry is shown as a thick gray line. The uncertainty range ($\pm$1$\sigma$) is shown in all cases. The extinction values for various lines of sight within the field based on the Besan\c{c}on Galactic model and 2MASS data \citep{Marshall06} are shown as filled dots ($A_{K_s}/A_V$ = 0.115 was assumed).}
\label{fig_extinction}
\end{figure}

As illustrated in Fig.\,\ref{fig_extinction}, the reddening toward the field LRc01 is ill-defined. In view of the relative uniformity of the extinction across the field and the large discrepancies between the various estimates \citep[see, e.g.][for a discussion of this problem]{Rowles09}, we adopt $A_V$ = 0.9 mag as a representative value for all stars in our sample. The extinction is predicted to rise with distance up to $\sim$1.2 kpc, but our targets are expected to lie beyond this point where it begins to taper off (Fig.\,\ref{fig_extinction}). The good agreement between the $T_{\rm eff}$ values derived from near-IR data and those determined from the optical photometric data in the Harris $BV$ and Sloan-Gunn $r^{\prime}i^{\prime}$ filters available in Exo-Dat suggests that this estimate is appropriate. The extinction in the $J$, $H$, and $K_S$ bandpasses were determined following \citet{Fiorucci03} assuming the spectral energy distribution of a K1 giant. 

\begin{figure}
\centering
\includegraphics[width=7cm]{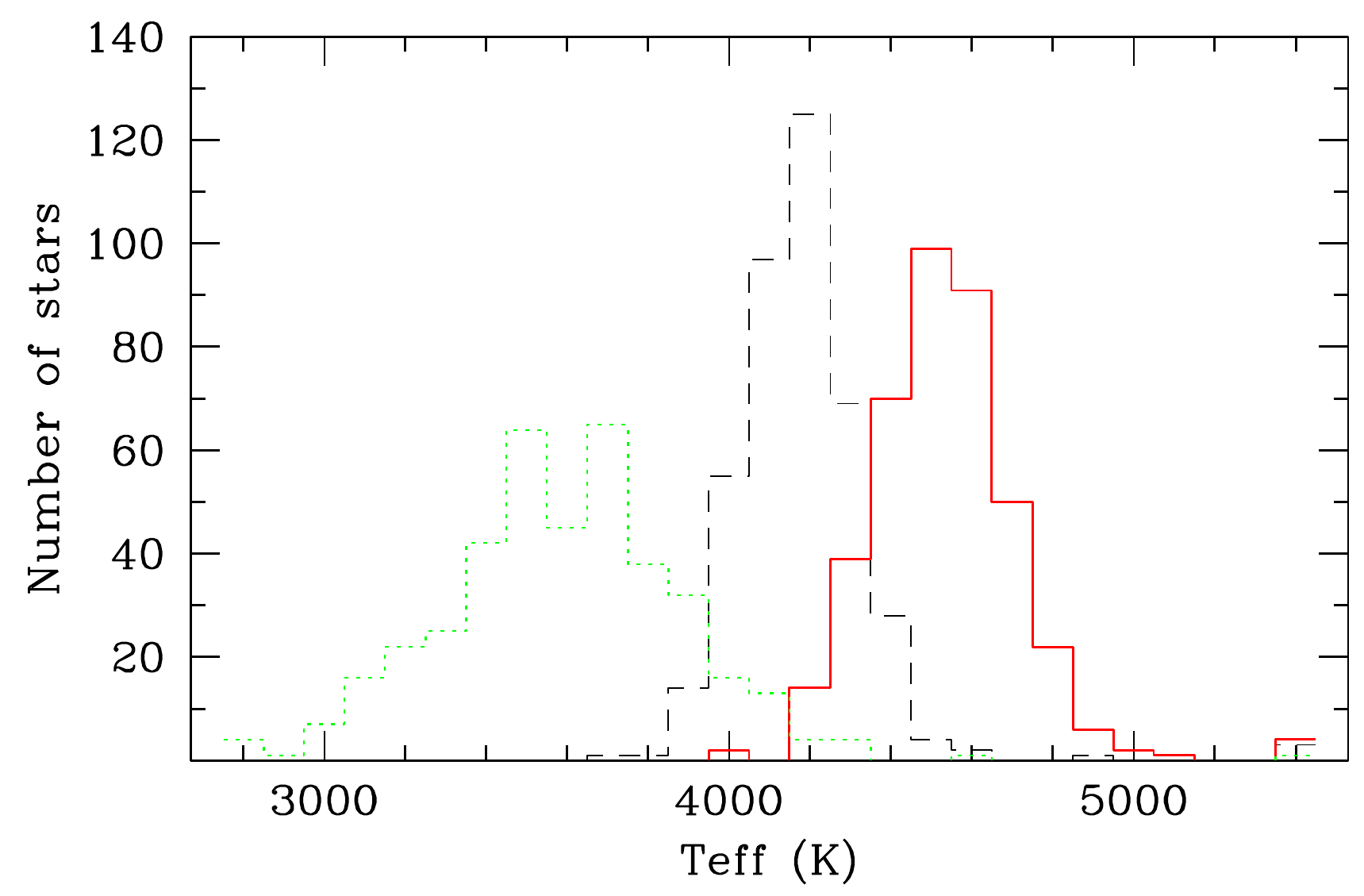}
\caption{Distribution of the effective temperatures before (black dashed line) and after dereddening (red full line) for the total sample of 398 stars. For comparison purposes, the values quoted in Exo-Dat are also shown (green dotted line). A typical value for the uncertainty is 150\,K.}
\label{fig_teff_distribution}
\end{figure}

We adopt the unweighted mean of $T_{\rm eff}$($J-H$) and $T_{\rm eff}$($J-K_S$) as the final $T_{\rm eff}$ values. A typical value for the uncertainty on these temperatures is $\sim$150\,K.
Their distribution, shown in Fig.\,\ref{fig_teff_distribution}, is centered on $\sim$4500K, which is about, but slightly lower than, the value expected for stars pertaining to the red clump, which are supposed to be representative of most of the red-giant stars observed here \citep{Miglio09}.

Another characteristic of interest is the luminosity-to-mass ratio, $L/M$. However, this ratio cannot generally be derived from observations in a straightforward and accurate manner. A scaling law can be derived to directly estimate $L/M$ from observables, namely $T_{\rm eff}$ and $\nu_{\rm max}$. First, we consider that $\nu_{\rm max}$, the frequency at which the modes have maximum amplitude, scales as the cut-off frequency $\nu_c$, which itself varies as $g/\sqrt{T_{\rm eff}}$ (where $g$ is the surface gravity). Then, as the $L/M$ ratio scales as $T_{\rm eff}^4/g$, we obtain
\begin{equation}
\frac{L}{M} \propto \frac{T_{\rm eff}^{7/2}}{\nu_{\rm max}}.
\label{eq:LM}
\end{equation}
This is only a scaling law and not a detailed description, but it is very useful to estimate the luminosity-mass ratio for a large number of red-giant stars exhibiting oscillations (and with a known $T_{\rm eff}$).

\subsection{Discussion}

Most of the red-giant stars that we investigated in this paper have been recently identified as belonging to the red clump \citep{Miglio09}. More precisely, these stars are post-flash helium-core burning stars. They are characterized by a $\nu_{\rm max}$ ranging from $20$ to $50 \, \mu$Hz, by masses between approximately $0.8$ and $2$ $M_\odot$, and by luminosities centered around $\log (L / L_{\odot}) \approx 1.75$.
Hence, the red giants observed by CoRoT are localized in a narrow interval range of mass and luminosity, which explains why the distributions in linewidth and amplitude are well centered, because most of the stars have similar physical properties.

Concerning the linewidths, the dichotomy between resolved and unresolved modes is difficult to interpret because it can be related to the presence of non-radial modes. A recent theoretical investigation of the power spectrum of red giants has been presented by \cite{Dupret09}. They consider different evolutionary stages for red giants and computed excitation and damping rates. The radial and non-radial modes trapped in the envelope can be detected with a 150-day long  observation. In terms of linewidth,  they are close to the resolution actually observed in Fig.\,\ref{fig:histolargamp}. From a theoretical point of view, it appears that radial modes are more likely to be resolved because they have a shorter lifetime \citep[see][for details]{Dupret09} than the non-radial modes. But it would be dangerous to identify resolved modes as radial and unresolved as non-radial, because the precise values of the damping rate critically depend on the convection-oscillation interaction, which is still poorly known.

The maximum amplitude plotted in Fig.\,\ref{fig:histolargamp} can also be qualitatively explained. Following \cite{Miglio09}, the red-clump CoRoT stars have radii ranging from approximately $10$ to $20 \, R_\odot$. In terms of effective temperature, they range roughly between $\log T_{\rm eff} = 3.7$ and $\log T_{\rm eff}=3.65$.
Mode amplitudes scale as \citep[see][for a detailed discussion of this relation]{Samadi07}
\begin{equation}
\label{eq:amp}
\left< A\right>_{\rm max} \propto \left(\frac{T_{\rm eff}^4}{g}\right)^{ s}
\propto \left(\frac{T_{\rm eff}^4 R^2}{M}\right)^{ s} .
\end{equation}
In addition, $\nu_{\rm max}$ has been shown to scale as the cut-off frequency $\nu_c$ \citep[see for example][]{Kjeldsen95}:
\begin{equation}
\label{eq:nu_max}
\nu_{\rm max} \propto \nu_c = \frac{c_s}{H_p} \propto \frac{M\, R^2}{\sqrt{T_{\rm eff}}},
\end{equation}
where $H_p$ is the pressure scale height, and $c_s$ the sound speed.
Then, from Eq.\,\ref{eq:nu_max}, the horizontal scatter seen in Fig.\,\ref{fig:amp_numax} can be expected from the dispersion in radius.
From Eqs.\,\ref{eq:amp} and \ref{eq:nu_max} the anti-correlation between the amplitude $A_{\rm max}$ and $\nu_{\rm max}$ is explained by
the dispersion in effective temperature of stars belonging to the red clump.
More precisely, Eq.\,\ref{eq:amp} shows that the higher $T_{\rm eff}$, the higher the mode amplitude, and Eq.\,\ref{eq:nu_max} shows that the higher $T_{\rm eff}$, the lower the $\nu_{\rm max}$.

\begin{figure}
\includegraphics[angle=90,height=7.5cm]{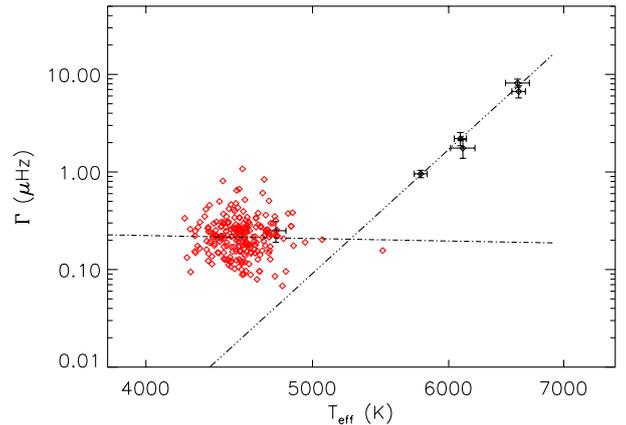}
\caption{Measured mode linewidths versus $T_{\rm eff}$ for the red giants ($T_{\rm eff} < 5000$\,K) and for the main-sequence stars ($T_{\rm eff} > 5000$\,K). Note the measured width for HD181907 ($T_{\rm eff}$=4760\,K) from \cite{Carrier10}. The dot-dashed line indicates the fit to the red-giant width, and the triple-dot-dashed line shows the fit to main-sequence stars.}
\label{fig:larg}
\end{figure}

\begin{figure}
\includegraphics[angle=90,height=7.5cm]{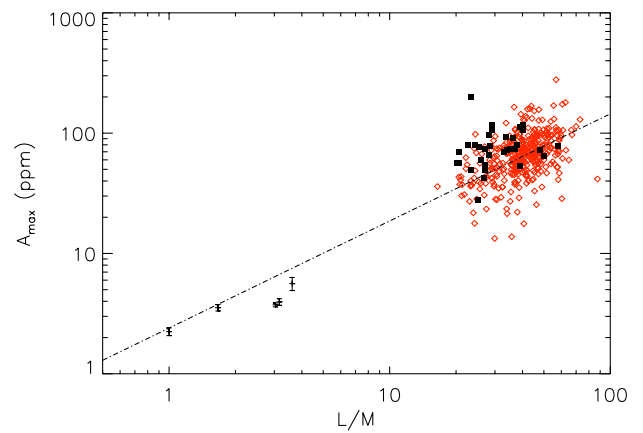}
\caption{Measured mode amplitudes ($\propto \sqrt{H \Gamma}$) versus $L/M$ for the red giants ($L/M > 5$) and for the main-sequence stars ($L/M < 5$). The results of  Barban et al. (in prep.) appear as filled squares. The result of \citet{Carrier10} for HD181907 is hidden by other red-giant results. The dot-dash line indicates the global fit to red giants and main-sequence stars.}
\label{fig:amax}
\end{figure}

In Fig.\,\ref{fig:larg} the variation of mode widths is shown versus the effective temperature of the stars. Apparently there is no link between temperature and mode width for the red giants. However, one must keep in mind the relatively large uncertainties on $T_{\rm eff}$ that can flatten the observed distribution.
However, when the amplitudes are shown versus the ratio $L/M$, there is a clear correlation. Given that $L/M$ is proportional to $T_{\rm eff}^{7/2}$ (see Eq.\,\ref{eq:LM}), this shows that the temperatures are clearly related with other observables despite their uncertainties.

\section{Comparison with other stars}
\label{sec:comp}

\subsection{Red-giant stars}

To check the validity of our results, the detailed results of \citet{Carrier10} for HD181907 were used, as well as those of Barban et al. (in prep.). HD181907 shows resolved modes, and the widths given in this paper are consistent with the widths measured here (see Fig.\,\ref{fig:larg}). Maximum mode amplitude is also similar to the ones derived here. Most (of 36) of the spectra fitted by Barban et al. (in prep.) show unresolved modes. Their fitted amplitudes show a distribution that agrees with our results (see Fig.\,\ref{fig:amax}).

\subsection{Main-sequence stars}

Red-giant stars observed by CoRoT are a unique opportunity to infer the seismic characteristics of these evolved stars. In addition, a comparison with main-sequence stars is also very useful to determine whether or not the pulsating properties of both groups of stars follow the same scaling laws.

Some main-sequence stars have shown solar-like oscillations. To use a homogeneous sample, we made our comparison using the recent results for solar-like pulsators observed by CoRoT: HD49933 \citep{Benomar09b}, HD181420 \citep{Barban09}, and HD52265 \citep[Ballot \& Benomar, private communication; ][]{Valenti05,Soriano07}. The results for 
HD49385 \citep{Deheuvels10,Deheuvels09b} were also considered, but this star has just left the main-sequence. The luminosities used in Table\,\ref{tab:HD} are taken from the above mentioned articles, as were the amplitudes and widths, which were derived from individual mode fitting and were averaged from the three highest radial orders. Error bars indicate the corresponding standard deviations.
In some other cases not considered here, oscillations were detected, but the characteristics of the individual modes are not available. This selection thus provides a homogeneous set of very high-quality measurements.
The stellar characteristics of all these stars are listed in Table\,\ref{tab:HD}. Their mass is derived from stellar modeling including seismic constraints, except for HD52265.\\
In addition to this group of stars, the Sun remains a reference. Data from the VIRGO/SPM instrument \citep{Frohlich97} were fitted and the mode width and amplitude derived in the same manner as for the other stars used here.\\

Because these stars span quite a large interval in effective temperature (from $\sim$4000\,K to $\sim$6500\,K), the bolometric correction \citep{Michel09} was also applied to the measured amplitudes (see Sect.\,\ref{sec:resRG}). Corrections are of the order of $\sim$10\%. 

\begin{table*}
\label{tab:MS}
\center{
 \begin{tabular}{|c|c|c|c|c|}
\hline
Star name & Amplitude (ppm) & Width ($\mu$Hz) & $T_{\rm eff}$ (K) & $(L/L_{\odot})/(M/M_{\odot})$ \\
\hline
HD49933 & 3. 43 $\pm$ 0.14 &  6.68 $\pm$ 0.56 & 6590 $\pm$ 60 & 2.8 \\
HD181420 & 3.63 $\pm$ 0.25 & 8.15 $\pm$ 0.78 & 6580 $\pm$ 105 & 3.3 \\
HD49385 & 5.45 $\pm$ 0.70 & 2.20 $\pm$ 0.2 & 6095 $\pm$ 65 & 3.7 \\
HD181907 & 104 $\pm$ 17 & 0.25 $\pm$ 0.06 & 4760 $\pm$ 65 & 58.3 \\
HD52265 & 3.43 $\pm$ 0.2 & 1.75 $\pm$ 0.37 & 6115 $\pm$ 100 & 1.5 \\
Sun & 2. 24 $\pm$ 0.16 & 0.95 $\pm$ 0.08 & 5780 $\pm$ 50 & 1 \\
\hline
\end{tabular}}
\caption{Stellar characteristics used in the comparison with our results. The masses used here are drawn from stellar modeling (including seismic constraints, except for HD52265) and not from scaling laws. A typical uncertainty on the $L/M$ ratio is 20\%.}
\label{tab:HD}
\end{table*}

\subsection{Scaling laws for mode linewidths}

No clear scaling law was found between widths and the ratio $L/M$ or with the cut-off frequency. We then compared the mode linewidths of the red giants (RG) with main-sequence (MS) stars versus effective temperature in Fig.\,\ref{fig:larg}.
The figure shows a linear fit (in the logarithm) for each group:
\begin{equation*}
\Gamma \propto T_{\rm eff}^s \quad \mbox{with} \quad\left\{
\begin{array}{ll}
s \approx -0.3 \pm 0.9\; &\text{for RG } \\
s \approx 16 \pm 2 \; &\text{for MS }.
\end{array} \right.
\end{equation*}
Despite the large uncertainties on these results, red giants and main-sequence stars show a different behavior: small linewidths with a flat distribution, compatible with no variation of width with $T_{\rm eff}$ for red giants, and a  strong dependence of the mode width with effective temperature for main-sequence stars (Fig.\,\ref{fig:larg}). However, the flat distribution for the red giants could caused by the relatively large uncertainties on their temperature. However, despite the large uncertainty on main-sequence width variation with temperature, it appears that a single power law 
is not sufficient to describe the mode damping over the whole range of $T_{\rm eff}$. It suggests that two different physical regimes occur for red giants and main-sequence stars. In the latter case, the mode widths show a very strong sensitivity to the effective temperature. We conclude that a theoretical computation of mode damping rates for red-giant stars cannot be calibrated from main-sequence stars (and thus from the Sun).

The accuracy of the CoRoT results shows a very good agreement of the widths measured with a scaling law for main-sequence stars. We emphasize again that the number of analyzed stars is small, but includes only stars for which individual mode fitting was possible. Our result contrasts with the results obtained by
\cite{Chaplin09}, who  found that the mode linewidth varies as $T_{\rm eff}^4$, based on the analysis of 12 solar-like main-sequence, sub-giant and giant stars mainly from ground-based observations and theoretical computations based on the \cite{Balmforth92,Houdek99,Chaplin05} formalisms. These ground-based observartions generally provide results with a large uncertainty that can be explained by the short duration and the discontinuous observation series and a inhomogeneous sample.

\subsection{Scaling law for mode amplitudes}

Because the results span a wide range of temperatures, the comparison of the mode amplitude from red giants to main-sequence stars requires a bolometric correction to be applied to the amplitudes, following \cite{Michel09}.
As mentioned in Sect.\,\ref{sec:intro}, previous work \citep[e.g., ][]{Houdek99,Houdek02,Samadi07} has used a variation of the maximum oscillation amplitude in terms of the ratio $(L/M)^s$ (with 0.7\,$<s<$\,1.5). We represent in Fig.\,\ref{fig:amax} the measured maximum mode amplitude versus this ratio, keeping in mind that  the results of \citet{Houdek02} and \citet{Samadi07} apply for  amplitudes seen in velocity, whereas we deal with intensity measurements.
As in the case of linewidth variations, a different scaling law is sought for each group of stars. The computed slopes (in logarithm) $s_{\rm MS}$ and $s_{\rm RG}$ are
\begin{equation*}
\label{eq:resamp}
A_{\rm max}^{\rm I} \propto \left(\frac{L}{M}\right)^s \quad \mbox{with} \quad\left\{
\begin{array}{ll}
s = 0.92 \pm 0.14 \; &\text{for RG } \\
s = 0.42 \pm 0.14 \; &\text{for MS },
\end{array} \right.
\end{equation*}
where $A_{\rm max}^{\rm I}$ is the observed maximum of amplitude in intensity.  

The slopes computed here are obtained from very different measurements: very few but very precise measurements for MS stars, and a large number of less precise measurements for RG stars. The uncertainty on $s_{RG}$ is more than two times better than on $s_{MS}$, thanks to the large number of stars used. This uncertainty was checked by computing the slope $s$ for RG stars from a sub-sample of the whole set (for example 80\%): the results obtained are consistent with the computed uncertainty from the whole set. The low number of MS stars used and the dispersion of their results induce a large uncertainty: despite a flatter slope, the difference with the slope measured for red-giant stars may not be significant. Thus, in this case, and contrary to the case of the mode widths, a unique scaling law was computed for all stars considered here. The global slope is \mbox{$s_{all}= 0.89 \pm 0.08$}. The corresponding power law is shown in Fig.\,\ref{fig:amax}. Even if the difference between the slopes $s_{RG}$ and $s_{MS}$ mentioned above may not be significant, all of the MS stars (except the Sun) fall below the global power law.

To derive a proper relation between amplitudes and $L/M$, one must take into account that the present amplitudes are drawn from intensity and not velocity. This can be done by considering the adiabatic case and using the relation between intensity and velocity as derived by \citep{Kjeldsen95}
\begin{equation}
\label{relation_iv_adia}
A_{\rm max}^{\rm I} \propto \frac{A_{\rm max}^{\rm v}}{\sqrt{T_{\rm eff}}},
\end{equation}
where $A_{\rm max}^{\rm v}$ is the maximum amplitude in velocity.
The slope computed using Eq.~\ref{relation_iv_adia} is not strongly affected: \mbox{$s_{all}^V= 0.86 \pm 0.08$}. This slope is comparable with the lower end of the previously reported interval of values [0.7-1.5]. However, one should not forget that the adiabatic relation may not be adequate to accurately describe the relation between mode amplitudes in intensity and velocity.

\section{Conclusion}

The first result that can be drawn from this global analysis of a large number of red-giant stars is that even if some dispersion exists, the large number of analyzed spectra allows the robust determination of general trends for this kind of stars, in particular about the variation of the maximum amplitude of the observed modes versus the frequency of these modes.
The widths of the modes are clearly narrow and of the same order as the frequency resolution of the Fourier spectra, which is very good however, coming from time series of more than 140 days. The observed dispersions in amplitudes and widths can have several origins. First, they can originate in the uncertainty of the automated fitting procedure. Second, when considering amplitude versus the luminosity-mass ratio, they can originate in the uncertainty in the determination of the effective temperature from which the luminosity-mass ratio is extracted. We recall here the importance of this temperature determination. Last, but not least, the dispersion can have an intrinsic stellar origin, because some stellar characteristics, such as the metallicity, are expected to have an influence on convection and thus on mode excitation \citep[see][]{Samadi10}.

The main results of the comparison of red giants with main-sequence stars are the scaling laws for mode parameters that can be drawn from this analysis, which spans large intervals of stellar parameters. In addition to the robustness of the results for red giants, the precision of the results from CoRoT observations of main-sequence stars make this kind of comparison fruitful. The mode linewidths in red giants are not, or very weakly, dependent on the effective temperature. On the contrary, they show a very strong dependence on $T_{\rm eff}$ for less evolved stars. The maximum mode amplitude dependence on the luminosity-mass ratio appears to be similar for RG and MS stars. To a first approximation, a single scaling law ranging from the Sun to red giants seems to describe the observations.

Thanks to the newly available asteroseismic measurements, it is now possible to provide precise constraints on the oscillation amplitude and damping across the H-R diagram. The widths measured for the modes of red giants are similar to those proposed by \citet{Dupret09} based on the strong link between mode inertia and mode linewidth. However, observations for a large number of stars and for a longer timespan than those used here would allow us to check the importance of the bias caused by the frequency resolution of the spectra used here. For the mode widths of main-sequence stars, our estimate of the exponent of the power law is based on a small number of stars, but the precision of the results indicates that the description proposed by \citet{Chaplin09} has to be reconsidered.

Concerning the maximum mode amplitudes, if considering a single scaling power law, its exponent ($\approx$\,0.8) is in the lower end of the range proposed in the literature \citep[0.7-1.5, see for example][]{Houdek99,Samadi07}. However, the main-sequence stars are all below this scaling law, which is also illustrated by the difference of a factor of two  when comparing the exponents computed for each group of stars. More MS stars would permit us to check if a unique scaling law describes both RG and MS stars or if, as for the mode width, two regimes are present. The observation of sub-giants would be particularly helpful for the description of the mode parameters and a possible transition regime between evolved and unevolved stars.

\begin{acknowledgements}
FB wishes to thank J. Leibacher for very helpful discussions.
This work made use of the Exo-Dat database, operated at LAM-OAMP, Marseilles, France, on behalf of the CoRoT/Exoplanet program. The research leading to these results has received funding from the European Research Council under the European Community's Seventh Framework Programme (FP7/2007--2013)/ERC grant agreement n$^\circ$227224 (PROSPERITY), as well as from the Research Council of K.U.Leuven (grant agreement GOA/2008/04). T.K. is supported by the Canadian Space Agency and the Austrian Science Fund. T.M. acknowledges financial support from Belspo for contract PRODEX-GAIA DPAC. J.B. acknowledges support through the ANR project Siroco. S.H. acknowledges financial support from the UK Science and Technology
Facilities Council (STFC).
\end{acknowledgements}

\bibliographystyle{aa}
\bibliography{RG}

\end{document}